\documentclass[12pt]{article}
\usepackage{a4wide}
\usepackage{amssymb}
\begin{document}

\newcommand{\uvec}[1]{\raisebox{-1.5mm}{$\stackrel{\textstyle #1}{\scriptscriptstyle\rightarrow}$}{}}

{\renewcommand{\thefootnote}{\fnsymbol{footnote}}
\hfill  IGC--11/9--1\\
\medskip
\begin{center}
{\LARGE Quantum gravity in the very early universe\footnote{Published as {\em Nuclear Physics} {\bf A862--863} (2011) 98--103}}\\
\vspace{1.5em}
Martin Bojowald\footnote{e-mail address: {\tt bojowald@gravity.psu.edu}}
\\
\vspace{0.5em}
Institute for Gravitation and the Cosmos,\\
The Pennsylvania State
University,\\
104 Davey Lab, University Park, PA 16802, USA\\
\vspace{1.5em}
\end{center}
}

\begin{abstract}
  General relativity describes the gravitational field geometrically
  and in a self-interacting way because it couples to all forms of
  energy, including its own. Both features make finding a quantum
  theory difficult, yet it is important in the high-energy regime of
  the very early universe.  This review article introduces some of the
  results for the quantum nature of space-time which indicate that
  there is a discrete, atomic picture not just for matter but also for
  space and time. At high energy scales, such deviations from the
  continuum affect the propagation of matter, the expansion of the
  universe, and perhaps even the form of symmetries such as Lorentz or
  CP transformations. All these effects may leave traces detectable by
  sensitive measurements, as pointed out here by examples.
\end{abstract}

Processes in the very early universe require for their description
general relativity (space is expanding) and quantum physics (the early
universe is hot and dense). Sometimes, this even involves quantum
physics not just of matter but of gravity. Gravity is described by the
geometry of space-time, and so we need to quantize space and time.  By
experience from quantum mechanics, one possible consequence is that
elementary constituents, or ``atoms of space,'' arise for space-time.

Dimensional arguments can be used to arrive at a first estimate of
direct effects. There is a unique length parameter, the Planck length
$\ell_{\rm Pl}=\sqrt{G\hbar/c^3}\approx 10^{-35}{\rm m}$ and a unique
mass parameter, the Planck mass $M_{\rm Pl}= \sqrt{\hbar c/G}\approx
10^{18}{\rm GeV}\approx 10^{-6}{\rm g}$, that can be formed solely by
reference to the relevant fundamental constants, Newton's
gravitational constant $G$, Planck's constant $\hbar$, and the speed
of light $c$.  At those scales, or, perhaps more intuitively, at the
Planck density $\rho_{\rm Pl}= M_{\rm Pl}/\ell_{\rm Pl}^3$, quantum
gravity becomes inevitable. Compared to the current density of the
universe, at about an atom per cubic meter, the Planck density of
roughly a trillion solar masses in the region of the size of a single
proton, is huge. The relevance of quantum gravity for current physics
may thus be questioned.

However, dimensional arguments can be misleading when large
dimensionless parameters are involved. In the context of quantum
gravity, perhaps suggesting some kind of elementary atoms of space,
such a parameter can easily be seen to arise: the number of tiny,
Planck-sized spatial atoms in a given macroscopic region under
consideration.  Similar questions, in which indirect evidence for
phenomena on tiny scales can be found long before the resolution of
observations becomes good enough for direct tests, have played
important roles before in the history of physics.  For instance, in
1905 Albert Einstein used an analysis of Brownian motion to find
convincing evidence for atoms, and only fifty years later, in 1955,
did Erwin M\"uller produce the first direct image of atoms using field
ion microscopy. By that time, the overwhelming majority of physicists
was already convinced of the reality of material atoms based on
Einstein's arguments.

Returning to quantum gravity, the best microscope we currently have to
magnify and probe the fundamental structures of space is the universe
itself. By its own expansion, it enlarges spatial regions and
eventually translates their properties into visible large-scale
structures. This magnification process, of course, takes a long time
and many other processes happen throughout; no direct image can be
obtained in this way and a great deal of physics must be used to
disentangle the form of original structures from what has emerged in
the meantime. With a good understanding of all the physics involved
one can begin to find indirect evidence for effects controlled by
quantum gravity.

The physics of quantum gravity is not well-understood at present, and
no complete theory is known. Nevertheless, several characteristic
effects have been suggested which do not so much depend on theoretical
details but are rather based on general expectations from fundamental
properties of general relativity and quantum mechanics. One of the
main such suggestions is the atomic nature of space, and it is of
relevance for early-universe cosmology. An expanding discrete space
grows not continuously but atom by atom.  Implications are weak for a
large universe, but may be noticeable by sensitive observations of
events that happen sufficiently early.

Observations which have a chance of providing sufficient sensitivity
with current technological means must first be found by analyzing
available theories. As an example one may consider the abundances of
light elements, which depend on the baryon-photon ratio during
big-bang nucleosynthesis, an early-universe process of
proton-neutron interconversion by the weak interaction.  The
baryon-photon ratio depends on the dilution behavior of radiation and
(relativistic) fermions. If discrete expansion leads to modifications
of the dilution behavior, small changes in the abundance of light
elements would be expected. We will return to this example at a later
stage.

Other examples for some chance of testing quantum gravity can be found
in all the phases included in the standard model of cosmology:
\begin{description}
\item[The big bang:] an extreme phase starting with Planckian density
  preceded, in the classical understanding of general relativity, by a
  singularity 13.8 billion years ago.
\item[Inflation:] an accelerated phase of expansion
of currently unknown origin, happening at an energy scale about
$10^{-10}\rho_{\rm Pl}$ at which particle production seeds all matter
as seen in the cosmic microwave
  background (CMB) and the galaxy distribution.
\item[Baryogenesis:] the formation of baryons out of a primordial
  quark-gluon plasma, somehow expected to lead to the
  matter/antimatter asymmetry of the present universe.
\item[Nucleosynthesis:] the generation of nuclei as bound states of
  the light baryons, arising in relative quantities of about 75\%
  hydrogen and deuterium, 25\% helium, and just trace amounts of other
  light elements.
\item[CMB release:] once atoms neutralize, the universe becomes
  transparent 380,000 years after the big bang.
\end{description}

The succession of most of these phases is well supported by
observations. However, the picture is incomplete, for the story begins
with a singularity at which the equations of general relativity lose
their meaning and unphysical conditions such as infinite densities and
temperatures are reached. The singularity is a general consequence of
the equations that govern a classical universe, in the simplest case
described by the Friedmann and Raychaudhuri equations
\begin{equation} \label{Friedmann}
 \left(\frac{\dot{a}}{a}\right)^2= \frac{8\pi G}{3}\rho\quad,\quad 
\frac{\ddot{a}}{a}= -\frac{4\pi G}{3} 
 (\rho+3P)
\end{equation}
for the scale factor $a$ providing the distance measure of the
universe, and with the energy density $\rho$ and pressure $P$ of
matter.

A simple singularity theorem can be obtained from this equation as
follows: First, a simple rewrite implies $ \dot{\cal H}= -4\pi G
(\frac{1}{3}\rho+P)-{\cal H}^2$ for the Hubble parameter ${\cal
H}=\dot{a}/a$.  If we assume the so-called strong energy condition
$\rho+3P\geq 0$ as a rather general requirement for the matter
ingredients, we obtain the inequality ${\rm d}{\cal H}^{-1}/{\rm
d}t\geq 1$ and thus ${\cal H}^{-1}\geq {\cal H}_0^{-1}+t-t_0$.  If
${\cal H}_0^{-1}$ is negative, ${\cal H}^{-1}$ must be positive at
$t_1=t_0-{\cal H}_0^{-1}$, and so ${\cal H}^{-1}=0$ at some time when
${\cal H}\to\infty$ and $\rho\to\infty$ diverge: a future singularity
resulting from collapse.  Similarly, a past singularity is obtained if
the universe is expanding at some time.  More complicated singularity
theorems can be demonstrated under more general conditions, dropping
the symmetry assumption of an exactly isotropic universe and weakening
the conditions posed for matter.  Thus, singularities are generic in
space-time dynamics.

These conclusions lead to the identification of several shortcomings
of the standard model of cosmology despite its observational success:
(i) Any singularity is unphysical and must be eliminated by
  improving the theory.
(ii) Inflation assumes that matter starts out in an initial vacuum
  state. Is this assumption appropriate, especially when the density
  and temperature diverge at the ``initial'' singularity?
(iii) With current theories, the matter/antimatter asymmetry that is
  supposed to form during baryogenesis is difficult to explain.  If
  there was a prehistory of the universe before the big bang, as one
  possible scenario alternative to a singular one, more time existed
  for an asymmetry to build up.
(iv) The matter equation of state is important for some aspects of
  big-bang and other phases, but is not well known for most of the
  scales between currently probed densities and the Planck
  density. This lack of knowledge does not so much affect the singular
  nature but plays a role for specific scenarios.
To see what ingredients exactly we must bring under better control to
discuss possible improvements of the standard model, we need more
information about quantum gravity and the resulting space-time
structure.

Gravity is ``strongly interacting'' at a fundamental, non-perturbative
level.  This statement may come as a surprise, given that gravity is
much weaker than the other fundamental forces and can safely be
ignored in particle interactions. However, particle physics already
provides indications for the special nature of gravity: Its well-known
non-renormalizability implies that gravity cannot be quantized as a
weakly-interacting theory of gravitons on some background space-time.
The weak form of gravity should rather arise as the long-range remnant
of a more elementary theory. What exactly this elementary theory is is
difficult to extract from the long-range physics of gravity that we
know. Theoretical models are based on suitable principles for their
mathematical formulation, for which different approaches exist, but no
fully consistent version yet.  

A quantization directly addressing the structure of space and time is
loop quantum gravity \cite{Rev}, based crucially on
the principle of background independence. Some part of the theory can
be constructed by means analogous to those of lattice QCD, but with
one crucial difference: General covariance implies that all states
must be invariant under deformations of space (diffeomorphisms or
coordinate changes). As a consequence, several new features (and
complications) compared to QCD arise:
(i) Regular lattices are too restrictive because they would be
  deformed when coordinates are changed. Instead lattices are
  ``floating;'' they are not assigned a fixed position in space. Only
  topological and combinatorial properties of their linking and
  knotting behavior can be relevant for gravity.
(ii) No well-motivated restriction on the valence of lattice vertices
  exists (except the desired but possibly deluding simplicity of their
  mathematical description).
(iii) Superpositions of different lattice states must be considered
  because the lattices correspond to states of a fundamental quantum
  theory, not to an approximation of such a theory.
(iv) States of the continuum theory are described by lattices; they
  do not provide an approximation, and no continuum limit is to be
  taken.
In this way, one obtains a fundamental lattice theory for quantum
geometry. Geometrical excitations are, as we will see, generated by 
creation operators
for lattice links. Near the continuum, physics can only be described
by a highly excited many-particle state; in this sense the theory is
``interacting''.  So far, the complicated resulting physics has mainly
been analyzed in model systems, primarily obtained by assuming spatial
symmetries.

To provide more technical details, we describe space-time geometry by
an su(2)-valued ``electric field'' $\vec{E}_i$ and a ``vector
potential'' $\uvec{A}_i$ (using so-called Ashtekar--Barbero variables)
with the following meaning.\footnote{In general relativity, we must
distinguish between contravariant and covariant vector fields, denoted
here by arrows above or below the letter, respectively. On a metric
manifold one can uniquely transform between these two types of vector
fields, but for gravity the metric follows from the fundamental
fields. It is not available before those fields are known. Keeping
track of the metric dependences is crucial for a
background-independent formulation of quantum gravity.}
\begin{description}
\item[Electric field:] Geometrically called a densitized triad, it
  determines spatial distances and angles by assigning three
  orthonormal vectors $\vec{E}_i$, $i=1,2,3$, to each point in space.
\item[Vector potential:] $\uvec{A}_i=\uvec{\Gamma}_i+\gamma
  \uvec{K}_i$ where $\uvec{\Gamma}_i$ is related to the intrinsic
  curvature of space, and $\uvec{K}_i$ to extrinsic curvature of space in
  space-time. These two contributions are added with a relative
  weighting $\gamma$, the real-valued Barbero--Immirzi parameter.
\end{description}

In addition to these geometrical properties and meanings of the
fields, we have their canonically conjugate nature: $\vec{E}_i$ is the
momentum of $\uvec{A}_i$, $\{\uvec{A}_i(x),\vec{E}_j(y)\}= 8\pi \gamma
G \delta_{ij} \,\vec{\uvec{\delta}} \, \delta(x,y)$ (using the
identity matrix $\vec{\uvec{\delta}}$).  We can thus proceed by
attempting a canonical quantization, with due observation of special
properties resulting from symmetries of the theory, in particular
general covariance.

As in lattice gauge theories, we define as basic variables holonomies
$h_e={\cal P}\exp(\frac{i}{2}\smallint_{e}{\rm d} \lambda
\uvec{A}_j\cdot\vec{t}_e \sigma^j)$ for the connection $\uvec{A}^i$
along spatial curves $e$, with Pauli matrices $\sigma^j$.  However, we
use these objects in a way very different from lattice gauge theory;
they will become creation operators of quantum geometry.  To that end,
we define a basic state $\psi_0$ by $\psi_0(\uvec{A}_i)=1$, that is it
is independent of the connection.\footnote{This state turns out to be
  normalizable by the inner product constructed via integration on
  spaces of connections \cite{FuncInt}.}  Excited states are then
obtained by the action of holonomies via multiplication in this
connection representation. We present the formulas only in a
simplified U(1)-example where $h_e(\uvec{A})= \exp(i\int_e{\rm d}
\lambda \uvec{A}\cdot\vec{t}_e)$ are just phase factors; SU(2)
formulas as needed for gravity are analogous but more tedious. We thus
write all excited states obtained in this way as
$\psi_{e_1,k_1;\ldots;e_i,k_i}= \hat{h}_{e_1}^{k_1}\cdots
\hat{h}_{e_i}^{k_i}\psi_0$.  A general state is then labeled by a
graph $g$, the collection of all curves used for holonomies to
generate the state, and integers $k_e$ as quantum numbers on the
edges: $\psi_{g,k}(\uvec{A})=\prod_{e\in g}
h_e(\uvec{A})^{k_e}=\prod_{e\in g} \exp(i k_e\smallint_e {\rm
  d}\lambda\uvec{A}\cdot \vec{t}_e)$.

The Ashtekar--Barbero connection has momenta $\vec{E}_i$ such that
$\sum_i\vec{E}_i\otimes \vec{E}_i= (\det \vec{\vec{\,q}})^{-1}\cdot
\vec{\vec{\,q}}$ gives the inverse spatial metric $\vec{\vec{\,q}}$.
Quantizing $\vec{E}_i$, or rather the fluxes $\int_S{\rm d}^2y
\uvec{n}\cdot\vec{E}_i$ (with $\uvec{n}$ the metric-independent
co-normal to surfaces $S$), they naturally become derivative
operators. At the level of states, flux operators measure the
excitation levels $k_e$:
\begin{equation} \label{Flux}
  \int_S{\rm d}^2y \uvec{n}\cdot\hat{\!\vec{E}}\psi_{g,k}=
\frac{8\pi\gamma G\hbar}{i}\int_S{\rm d}^2y \uvec{n}\cdot \frac{\delta
  \psi_{g,k}}{\delta \uvec{A}(y)}
=8\pi\gamma \ell_{\rm Pl}^2\sum_{e\in g} k_e{\rm  Int}(S,e) \psi_{g,k}
\end{equation}
with the intersection number ${\rm Int}(S,e)$. From this equation one
readily concludes that the $\psi_{g,k}$ are eigenstates of fluxes,
with eigenvalues given by $8\pi\gamma \ell_{\rm Pl}^2$ times an
integer.  Spatial geometry is discrete: for gravity, fluxes
representing the spatial metric have discrete spectra, and so do
operators for area or volume constructed from them \cite{AreaVol}. The
Planck length $\ell_{\rm Pl}=\sqrt{G\hbar}$ together with the
Barbero--Immirzi parameter determines the elementary discreteness
scale. From computations of black-hole entropy one derives that
$\gamma$ is of the order one, but somewhat smaller than one
\cite{Gamma2}.

So far, the quantum geometry we developed is only of space, not
space-time. In order to see how the graph states evolve in time,
possibly being reconnected and refined by the creation of new
vertices, we need to quantize the Hamiltonian. Schematically, it has
the form \cite{QSDI} $\hat{H}\psi_{g,k}=
\sum_{v,IJK}\epsilon^{IJK}{\rm tr}(h_{v,e_I}h_{v+e_I,e_J}
h_{v+e_J,e_I}^{-1} h_{v,e_J}^{-1} h_{v,e_K}
[h_{v,e_K}^{-1},\hat{V}])\psi_{g,k}$ summing over vertices $v$ of the
graph $g$ and triples $(IJK)$ of edges. As is clear from this
expression, there are creation operators (holonomies) as well as the
volume operator $\hat{V}$. At this fundamental level of elementary
excitations of geometry, the theory is interacting, as promised: The
Hamiltonian contains products of creation operators. Its action
describes the dynamics of a discrete graph state, depending on the
spatial geometry via the volume operator.

Gauge fields for fundamental forces other than gravity can be
implemented in a very similar way, via independent types of holonomies
for connections associated with the groups of the standard model.
Fermions are represented as spinor degrees of freedom in the vertices
of graphs, and the resulting matter Hamiltonian is added to $\hat{H}$
to form the total Hamiltonian. It turns out to be well-defined,
without divergences \cite{QSDV}.  However, the limit of a classical
space-time remains poorly understood, and so it is difficult to say
how exactly the theory breaks the bad spell of non-renormalizability.

Another main challenge remains, that of understanding space-time
dynamics. The elementary interactions mediated by holonomies as
creation operators must somehow conspire to result in the well-known
long-range behavior of gravity. In a cosmological context, for
instance, an expanding universe must result from the single tiny
bricks of Planck cubes added on to space when holonomies act. The task
is especially difficult owing to strong consistency requirements to
ensure that discrete quantum geometry combines in the correct way with
general covariance, normally thought of as a continuous group of
space-time transformations that cannot leave discrete structures
invariant.

Irrespective of the precise form of a consistent Hamiltonian, general
properties of the dynamics, required to implement background
independence, directly lead to several characteristic types of
quantum corrections.
\begin{description}
\item[Inverse-volume corrections:] Inverse metric components are
  corrected in all Hamiltonians because flux operators (\ref{Flux})
  have discrete spectra containing zero. Such operators do not have a
  densely defined inverse, which would be needed to quantize
  components of the inverse densitized triad. Operators with those
  inverses as their classical limit can be defined \cite{QSDV}, but
  since they are not direct inverse operators they imply quantum
  corrections at small flux eigenvalues.
\item[Holonomy corrections:] Higher powers of curvature components
  arise from the substitution of connection components by holonomies.
  Covariant representations must also include higher time derivatives,
  which come from the last type of corrections.
\item[Quantum back-reaction:] As always in interacting quantum
  theories, the evolution of expectation values depends on the
  behavior of the whole state, for instance on the development of
  fluctuations or correlations. This quantum interaction can be
  captured in canonical effective equations \cite{EffAc}.
\end{description}
While the first two types of corrections are characteristic of loop
quantum gravity, the third one is generic for all kinds of interacting
quantum systems.  One can illustrate some features of those
corrections, or more generally the dynamics of loop quantum gravity,
by considering reduced models. If one requires isotropy of space, the
key loop properties turn out to be preserved, but the dynamics
obtained from the same constructions is much easier to analyze. In the
resulting loop quantum cosmology \cite{LQC}, the Friedmann equation
(\ref{Friedmann}) receives corrections by higher powers of the
momentum $p_a$, the isotropic reduction of the connection, because it
is, according to the basic premise of loop quantization, replaced by
$\sin(\delta p_a)/\delta$ with some parameter $\delta$ akin to the
edge length in holonomies.


Such a whole series of higher-order corrections seems difficult to
control, but there is a special solvable model for matter given by a
free, massless scalar, in which the series can be resummed to change
(\ref{Friedmann}) to $\left(\dot{a}/a\right)^2= (8\pi G/3)\rho\left(1-
  \rho/\rho_0\right)$ \cite{RSLoopDual} with $\rho_0$ of the order of
$\rho_{\rm Pl}$.  (Solvability is based on an underlying ${\rm
  sl}(2,{\mathbb R})$ symmetry obtained from the algebra
$[\hat{V},\hat{J}]=i\delta\hbar\hat{H}$, 
$[\hat{V},\hat{H}]=-i\delta\hbar\hat{J}$,
$[\hat{J},\hat{H}]=i\delta\hbar\hat{V}$ with the volume $\hat{V}$,
$J=V\exp(i\delta{\cal H})$ for the Hubble parameter ${\cal H}$, and the
Hamiltonian $\hat{H}$ \cite{BouncePert}.)


Exact solutions show some of the main implications, intuitively
grasped as follows.  In a discrete space, as it underlies the
construction of this model, there is a finite capacity to store
energy. When the limit is reached, gravity turns repulsive at high
densities.  A bounce at about the Planck density results which, if it
is realized generally enough, can resolve the singularity problem.  At
present it remains unclear how general the mechanism to resolve
singularities is; it applies at least in models in which the kinetic
energy of matter dominates the potential term. It also remains to be
determined at what density $\rho_0$ the bounce happens. The precise
value depends on $\delta$ which is difficult to derive from the full
Hamiltonian. However, there are internal consistency conditions by
comparing the different corrections within the model, and those
conditions suggest that the bounce density is less than Planckian
\cite{Consistent}.

An interesting consequence is that matter properties are relevant
throughout cosmic evolution, including the bounce phase. Sometimes,
one attempts to develop this bounce cosmology \cite{BounceReview} as
an alternative to inflation to explain the nearly scale-free spectrum
of anisotropies.  In some models, structure can be generated in the
collapse phase and transmitted through the bounce. The transmission
phase is difficult to control because it is very sensitive to
quantum-gravity properties. Non-standard equations of state of exotic
matter leave an imprint on the structures formed.  Exotic matter may
also play a role in the build-up of anisotropy.

We now return to the example of the sensitive phase of big-bang
nucleosynthesis. In quantum gravity the Maxwell and Dirac Hamiltonians
could be subject to different quantum corrections, and thus the
relative dilution behavior may change.  Only inverse-triad corrections
have been implemented so far, which change the equations of state in
the same way for photons and relativistic fermions
\cite{FermionHolst}. Effects are thus not as strong as could have been
expected, but they are nevertheless close to being interesting: A
detailed analysis \cite{FermionBBN} provides an upper bound
$\rho<3/\ell_{\rm Pl}^3$ for the density of atoms of space, not off by
orders of magnitude from the theoretical expectation of at most about
one atom per Planck cube.  This looks promising; however, the
precision of big-bang nucleosynthesis observations at the present
stage is difficult to improve.  There is more potential in looking at
details of the cosmic microwave background. Here, Hamiltonians are
endowed with correction factors $\alpha\sim 1+\epsilon$ for
inverse-triad components in loop quantum gravity.  The parameter
$\epsilon$ can be constrained by a CMB analysis, and so far is
consistent with zero. However, there is a convergence of 
theoretical lower and observational upper bounds for the
parameters \cite{LoopRunning} which should accelerate with new data.


Another test area of quantum gravity is black holes. In general
relativity it is impossible, under very general assumptions on the
equation of state, to stop the gravitational collapse of a heavy star.
Gravity is always attractive, and thus becomes the dominant force when
matter is sufficiently dense.  In quantum gravity, the space-time
dynamics changes, and as in the solvable cosmological model we have
repulsive gravity at extremely high densities. Also for black holes, a
non-singular collapse results, but one that still leads to a horizon
trapping light \cite{BHPara}. However, the horizon disappears once the
collapsing matter has traversed the high-density phase. Horizons, and
thus by definition black holes, exist only for finite times.  The
horizon shrinks by Hawking evaporation, and eventually disappears, at
which time one expects some kind of stellar explosion. Also here,
specific models for collapse depend on the matter behavior, opening
ways for tests.

The space-time structure in quantum gravity may even have implications
for particle physics, especially for parity symmetry
\cite{FermionHolst}. The vector potential is defined as
$\uvec{A}_i=\uvec{\Gamma}_i+\gamma \uvec{K}_i$ where $\uvec{\Gamma}_i$
is parity-odd and $\uvec{K}_i$ parity-even.  Unless $\gamma$ is a
pseudoscalar, which for a fundamental constant would be rather
unusual, there is a non-trivial and indefinite parity behavior of
$\uvec{A}_i$.  Classically the equations of motion are parity
invariant; they are, after all, equivalent to Einstein's equation.
But there is so far no good reason to expect the invariance to be
preserved after replacing $\uvec{A}_i$ with $h_e(\uvec{A}_i)$,
implementing one form of quantum corrections. Parity is still to be
checked by involved calculations that not only derive corrections in
equations of motion but also ensure that they are consistent in the
sense of covariance and anomaly freedom.  If parity violation due to
quantum gravity is found, it may be relevant for baryogenesis.  In
this context, it is also worth mentioning that some bounce models show
a change of orientation at the densest moment (the universe ``turns
its inside out'').  Parity breaking will then become relevant for the
big-bang transition.

In order to discuss possible relationships between quantum
gravity and the quark-gluon plasma, the main topic of these
proceedings, the first thing to note is that there are still many
orders of magnitude from quark-gluon plasma densities to the Planck
scale. At best, indirect consequences can be expected as always in
quantum gravity. The following suggestions can be made:
\begin{itemize}
\item The matter equation of state is important for collapse and
  bounce scenarios as exemplified in here, for instance for the
  build-up of anisotropy and the evolution of structure.
\item The cosmological prehistory is relevant for baryogenesis: is it
  more reasonable to assume a matter/antimatter-symmetric initial
  state, or a more messy and non-symmetric one after the collapse
  of an entire universe?
 \item There are indications that symmetries such as parity or local
 Lorentz transformations are modified by quantum geometry, with
 implications for quantum field theory.
\end{itemize}

To summarize, we have considered a quantum theory of space-time as a
gauge theory. A crucial new feature compared to other gauge theories
is the important role of general covariance.  In loop quantum gravity,
this is seen to imply an (irregular) lattice structure even for the
continuum theory.  Direct effects are important only at extremely high
densities, but indirect tests are conceivable in intermediate regimes;
several examples have already been described in cosmology. For
specific scenarios, the equation of state of matter is then required
for details. There is certainly no observation yet or in the
foreseeable future, but bounds on the theory are becoming interesting
and have already ruled out some possibilities.

\section*{Acknowledgements} 

Part of this work was supported by
NSF grant PHY0748336. The author is grateful to the organizers of the
``6th International Conference on Physics and Astrophysics of Quark
Gluon Plasma'' (ICPAQGP 2010), where this work has been presented as a
plenary talk.


\end{document}